\documentstyle[11pt]{article}

	\newcommand{\DWsection}[1]{
		\refstepcounter{section}
		\section*{\large\bf{\thesection\ \ #1}}
		\setcounter{equation}{0} }
	\renewcommand{\theequation}{\thesection.\arabic{equation}}

	\newcommand{\DWsubsectionstar}[1]{ \subsection*{{\normalsize\bf #1}} }

	\newcommand{\ds}{\displaystyle}
	\newcommand{\tb}{& \displaystyle}
	\newcommand{\nl}{\\ \displaystyle}
	\newlength{\eqlength}
	\newcommand{\eq}{& \makebox[\eqlength]{=} &}

	\newenvironment{eqn}
		{
		\[  \begin{minipage}{\textwidth}
		\vspace{-\abovedisplayskip}
		\begin{eqnarray*}
		}{
		\end{eqnarray*}  \end{minipage}
		\refstepcounter{equation} \makebox[0pt][r]{(\theequation)}  \]
		}

	\newcounter{storeeqn}
	\renewcommand{\thestoreeqn}{\thesection.\arabic{storeeqn}}
	\newenvironment{num}
		{
		\setcounter{storeeqn}{\value{equation}}  \setcounter{equation}{0}
		\refstepcounter{storeeqn}
		\renewcommand{\theequation}{\thestoreeqn\alph{equation}}
		}{
		\setcounter{equation}{\value{storeeqn}}
		}

	\newcommand{\cen}[1]{\tb \makebox[-4\arraycolsep]{$#1$} &}

	\newcommand{\smfr}[2]{{\textstyle \frac{#1}{#2}}}

\begin{document}

\begin{titlepage}

\renewcommand{\thefootnote}{\fnsymbol{footnote}}

{ \large
	\hspace*{\fill}  EFI-92-43 \\
	\hspace*{\fill}  hep-th/9210031 \\
	\hspace*{\fill} August 1992 }

\vspace{1.0in}

\begin{center}
	{ \LARGE \bf Charged String-like Solutions of Low-energy Heterotic
	String Theory}
\end{center}
\vspace{0.15in}
\begin{center}
	{\large Daniel Waldram \protect\footnotemark[1] \\ }
	\medskip
	{\large \it Enrico Fermi Institute \\
		University of Chicago \\ 5640 South Ellis Avenue \\
		Chicago Illinois 60637 USA }
\end{center}

\vspace{0.3in}

\begin{center}  {\large \bf Abstract}  \end{center}
\bigskip
Two string-like solutions to the equations of motion of the low-energy
effective action for the heterotic string are found, each a source of
electric and magnetic fields. The first carries an electric current equal
to the electric charge per unit length and is the most general solution
which preserves one half of the supersymmetries. The second is the most
general charged solution with an event horizon, a `black string'. The
relationship of the solutions to fundamental, macroscopic heterotic strings
is discussed, and in particular it is shown that any stable state of such a
fundamental string also preserves one half of the supersymmetries, in the
same manner as the first solution.

\footnotetext[1] {e-mail: {\tt waldram@yukawa.uchicago.edu}}
\renewcommand{\thefootnote}{\arabic{footnote}}

\end{titlepage}

\DWsection{Introduction}					\label{introduction}

In the hope of understanding more about the structure of string theory by
studying non-perturbative results, some interest has been focused recently
on classical solutions to the theory which have the properties of solitons
or instantons. Starting with the low-energy effective action for the
massless, bosonic modes of the heterotic string, Dabholkar {\it
et~al.}~\cite{nonrenorm,sustrs & solitons} found a singular string-like
solution sharing many of the properties of supersymmetric solitons.
In some sense, the solution represents the fields around a macroscopic
fundamental string, since the singularities can be matched onto source
terms from a sigma-model action describing the coupling of the fundamental
string worldsheet to the spacetime fields. Starting with the same action,
though restricted to a ten-dimensional spacetime,
Strominger~\cite{heterotic solitons} found a supersymmetric five-brane
solution, which is essentially a Yang-Mills instanton in the four
transverse dimensions, and so is a true soliton without singularities. He
also conjectured that five-branes might be dual to heterotic strings, in
the sense of Montenon and Olive~\cite{duality}, that string theory has two
equivalent, `dual' formulations, one a theory of strings where the
five-branes appear as solitons, and one a theory of five-branes with the
strings as solitons, and where weak coupling in one formulation corresponds
to strong coupling in the other.

Giving a different perspective, Horowitz and Strominger~\cite{black strs}
have found that the solution of Dabholkar {\em et~al.}~\cite{sustrs &
solitons} is the extremal limit of a general `black string' solution (one
with an event horizon), suggesting that these solutions may prove to be
useful models for studying black holes (or in general black $p$-branes) in
the context of string theory.

The supersymmetric solution of Dabholkar {\em et~al.} and the black string
of Horowitz and Strominger both set the gauge fields in the low-energy
effective action to zero. The purpose of this paper is to generalize these
two solutions to include the presence of a non-zero $U(1)$ gauge field, and
so find the most general {\em charged} string-like solutions in each case.

The paper is divided as follows. Section~\ref{susy stuff} describes, in
terms of the usual first-quantized theory, the supersymmetries of a
macroscopic heterotic string. It emerges that any stable state is invariant
under half of the supersymmetries, a property which should be reflected in
any string-like solution representing the fields around such a fundamental
string. This constraint is then used in section~\ref{a solution} to find
the most general charged version of the solution of Dabholkar {\it et~al.}.
The parameters of this solution are interpreted in terms of the properties
of a fundamental string and a multiple-string solution is also given.
Section~\ref{another solution} gives the most general charged black string
solution, identifying both the extremal supersymmetric limit and the limit
giving the uncharged solution of Horowitz and Strominger.
Section~\ref{conclusion} briefly summarizes the results.

\DWsection{The supersymmetries of macroscopic heterotic strings}
											\label{susy stuff}

It is a common property of most of the soliton solutions of supersymmetric
theories that the solution preserves one half of the supersymmetries and so
saturates a Bogomol'nyi bound~\cite{susy str solitons}. The broken
supersymmetries generate at least some of the fermionic zero modes of the
soliton. This property is useful when looking for soliton solutions since
the equations for the supersymmetric variations of the fields are only
first order, whereas the equations of motion are second order. For the
solution of~\cite{sustrs & solitons} in ten dimensions, the supersymmetry
transformations are described by a Majorana-Weyl spinor $\epsilon$, and
which supersymmetries are broken by the solution depends on the chirality
of $\epsilon$ in the eight-dimensional space transverse to the static
string-like solution. Should the supersymmetries split in this way for a
general string-like solution in ten dimensions?

The string-like solutions to the low-energy field theory for the heterotic
string are taken to represent, at least to first order in the string
coupling constant $\alpha'$, the fields around a stable macroscopic
heterotic string. Consequently the symmetries of the solution should be
reflected in the symmetries of the fundamental, first-quantized string.

Fundamental macroscopic strings in $D$-dimensional spacetime are described
in~\cite{sustrs & solitons}. Heterotic strings exist in ten dimensions so
first $10-D$ dimensions are compactified and replaced by a superconformal
field theory. Then to construct a macroscopic string, one of the remaining
spatial dimensions is also compactified onto a circle of large radius.
Strings with non-zero winding number about this direction are designated as
macroscopic states, since the string must be long enough to loop around the
circle at least once. A tower of stable, purely bosonic, macroscopic states
of increasing winding number can be constructed out of the product of a
ground state of the right-moving, supersymmetric sector and various excited
states from the left-moving, gauge sector.

For $D=10$ these stable states can be described in the light-cone gauge
following~\cite{GSW:GS sustr}. Spacetime is considered as a product of a
two-dimensional Minkowski space and an eight-dimensional transverse
Euclidean space. Ten-dimensional Majorana-Weyl spinors, necessarily of a
definite chirality, say $+1$, then decompose under $SO(9,1) \supset SO(1,1)
\otimes SO(8)$ so that
	\begin{equation}						\label{spinor decomp}
		{\bf 16} \to \left( {\bf 1}_+,{\bf 8}_+ \right)
			\oplus \left( {\bf 1}_-,{\bf 8}_- \right),
	\end{equation}
where the subscripts represent the spinor's chirality in the subspaces. As
ever the left- and right-moving modes of the string fully decouple, and for
the latter, considered as comprising half a Green-Schwarz superstring, by a
judicious use of the $\kappa$- and residual conformal symmetry, the bosonic
and fermionic coordinates can be made to satisfy free wave equations and
decomposed into Fourier modes $\alpha^i_n$ and $S^a_n$ respectively. (Here
$i$ refers to a vector index in the transverse $SO(8)$ space, $a$ to a
${\bf 8}_+$ spinor index and shortly ${\dot a}$ will refer to a ${\bf 8}_-$
spinor index; $n$ labels the mode.) The supersymmetry charge $Q$ transforms
as a Majorana-Weyl spinor, so for a general infinitesimal transformation
parametrized by a Majorana-Weyl spinor $\epsilon=(\zeta^a,\eta^{\dot a})$
the operator generating the field variations decomposes, so that
	\begin{equation}						\label{sucharge decomp}
		\epsilon Q =  \zeta^a Q^a + \eta^{\dot a} Q^{\dot a},
	\end{equation}
where the charges have the mode expansions
	\begin{eqn}								\label{general sucharge}
		\tb
			Q^a = (2p^+)^{1/2} {S^a_0},
		& \\ \tb
			Q^{\dot a} = (p^+)^{-1/2} {\gamma_{\dot a a}^i}
				\sum \limits_{-\infty }^\infty {S_{-n}^a} {\alpha _n^i}.
	\end{eqn}
The coefficient $p^+$ is one of the light-cone momenta, $p^\pm = \left( p^0
\pm p^1 \right)/{\sqrt2}$, where the superscripts refer to orthogonal time-
and space-like directions in the two-dimensional Minkowski space, whilst
the coefficients $\gamma_{\dot a a}^i$ are the building blocks for a set of
real eight-dimensional Dirac matrices, or alternatively the Clebsch-Gordon
coefficients relating the vector ($i$ index) and two spinor ($a$ and $\dot
a$ indices) representations of $SO(8)$.

How does the supersymmetry act on the stable macroscopic states? First
recall that these states were ground states of the right-moving sector, and
further that by definition any annihilation operator --- a mode with $n<0$
--- acting on a ground state gives zero. Thus since $\alpha _0^i$ is simply
the transverse momentum $p^i$, the supersymmetry generator acting on a
right-moving ground state $\left| \phi_0 \right>$ becomes a linear
combination of the fermionic zero mode operators $S_0^a$, namely
	\begin{equation}						\label{sucharge on gs}
		{\epsilon Q} {\left| \phi_0 \right>}
		= {1 \over \sqrt{p^+}}
			\left( {\sqrt2} {p^+} {\zeta^a}
				+ {p^i} {\eta^{\dot a}} {\gamma_{{\dot a} a}^i} \right)
			{S_0^a} {\left| \phi_0 \right>}.
	\end{equation}

It would be natural to choose coordinates so that the string is at rest in
the transverse space, that is, to choose one of the frames where $p^i=0$,
so that all dependence on $\eta^{\dot a}$ in~(\ref{sucharge on gs}) then
disappears, implying that half the supersymmetries vanish. However since
the ground state of the supersymmetric sector has $p^2=0$, then a frame
where $p^i=0$ has either $p^+=0$ or $p^-=0$ and the light-cone gauge can
not then be used.

Nevertheless, it is still possible to write the supersymmmetry generator in
terms of spinors of definite chirality in a transverse space in which the
string is at rest. For instance, consider first some general frame used for
the light-cone gauge. Clearly, simply making a Lorentz boost along $p^i$ to
`catch up' with the string gives a second frame with the string at rest in
the transverse space. The spinor $\epsilon$ can be decomposed into parts of
opposite chirality in the transverse space of the second frame, labelled
$\zeta'$ and $\eta'$, so that in the first frame
	\begin{equation}						\label{boosted spinor}
		\epsilon = \left(
			\begin{array}{c}\ds
				{\zeta^a} \nl {\eta^{\dot a}}
			\end{array} \right)
		= {1 \over \sqrt{ {p^+} \left({p^+}-{p^-}\right)}} \left(
			\begin{array}{cc}\ds
				p^+  \tb  -{p^i}{\gamma_{\dot a a}^i}/\sqrt{2} \nl
				-{p^i}{\gamma_{\dot a a}^i}/\sqrt{2}  \tb  p^+
			\end{array} \right)
			\left( \begin{array}{c}\ds
				{\zeta'^a} \nl {\eta'^{\dot a}}
			\end{array} \right)
		= U^{-1} \left(
			\begin{array}{c}\ds
				{\zeta'^a} \nl {\eta'^{\dot a}}
			\end{array} \right),
	\end{equation}
where $U$ is the Lorentz boost along $p^i$. Furthermore the supersymmetry
generator acting on the ground state~(\ref{sucharge on gs}) can be written
as
	\begin{equation}						\label{boosted sucharge}
		{Q}{\left| \phi_0 \right>} = \sqrt{2({p^+}-{p^-})} {\,} {U}
			\left( \begin{array}{c} \ds {S_0^a} \nl 0 \end{array} \right)
			{\left| \phi_0 \right>}
	\end{equation}
so that, with~${}^{\rm T}$ representing transposition,
	\begin{equation}						\label{half sucharge}
		{\epsilon Q}{\left| \phi_0 \right>} = \sqrt{2({p^+}-{p^-})}
			\left(
				\begin{array}{c}\ds
					{\zeta'^a} \nl {\eta'^{\dot a}}
				\end{array} \right)^{\!\rm T}
			{\!} {U^{-1}} {U}
			\left(
				\begin{array}{c}\ds
					{S_0^a} \nl 0
				\end{array} \right)
			{\left| \phi_0 \right>}
		= \sqrt{2({p^+}-{p^-})} {\,} {\zeta'^a} {S_0^a}
			{\left| \phi_0 \right>}
	\end{equation}
is independent of $\eta'^{\dot a}$. The choice of frame for the light-cone
gauge was quite arbitrary, so the conclusion is that for any 8-dimensional
Euclidean subspace in which the string is at rest, if $\epsilon$ is of
negative chirality in this subspace and so can be parametrized by
$\eta'^{\dot a}$, then the action of ${\epsilon}Q$ on the ground state is
zero, and the ground state is invariant under these supersymmetries.
However, if $\epsilon$ is of positive chirality in the subspace, and so is
parametrized by $\zeta'^a$, the action of $\epsilon Q$ depends on the
action of $S_0^a$ and so is generally non-zero, and these supersymmetries
are broken. For a macroscopic string it is natural to choose the subspace
by first boosting to a frame where the string has motion only along its own
length and then taking the space orthogonal to the string worldsheet.

In summary, it is indeed a generic property of stable macroscopic strings,
or more generally of any state of the heterotic string which is a ground
state of the right-moving sector, that half the supersymmetries are broken,
and half preserved, depending on the chirality of the supersymmetry
parameter $\epsilon$ in the space transverse to the string worldsheet in
which the string is at rest. Any string-like soliton purported to
approximately represent the fields around such a stable macroscopic string
should reflect these symmetries.

\DWsection{A charged supersymmetric solution}
											\label{a solution}

This section describes a new string-like solution to the low-energy
effective action for the massless modes of the heterotic string which
includes a $U(1)$ gauge field. Starting with a suitable ansatz, a partial
solution is obtained by imposing the condition that half the
supersymmetries are broken. The solution is completed using the equations
of motion.

To first order in the string loop expansion, the effective action is simply
that of ten-dimensional supergravity coupled to super Yang-Mills. For the
bosonic modes, namely the graviton $g_{\mu\nu}$, the antisymmetric tensor
field $B_{\mu\nu}$, the dilaton $\phi$ and the gauge potential $A_\mu$,
generalized to $D$ dimensions, it is given by
	\begin{equation}						\label{effective action}
		S = {1\over2\kappa^2}
			\int_{}^{} {{\rm d}^D{\!}x}{\,} {\sqrt{-g}}{\,}
				{{\rm e}^{-\phi/\alpha}}
				\left(
					R + \left({\partial \left(\phi/\alpha\right)}\right)^2
					- {\smfr{1}{3}} {H^2}
					- {\frac{\alpha'}{30}} {\rm tr}{F^2}
					\right),
	\end{equation}
where $\alpha=\sqrt{2/(D-2)}$, $\alpha'$ is the string coupling constant
and $\kappa$ is the gravitational coupling constant. The notation follows
\cite{heterotic solitons} in using the so-called `sigma-model metric' which
arises naturally when the action is derived by considering a string moving
in a classical background of the massless bosonic fields.
The physical metric $\tilde{g}$, for which the action has the usual
curvature term without the additional ${\rm e}^{-\phi/\alpha}$ factor, is
related by ${\tilde g}_{\mu\nu}={\rm e}^{-\alpha\phi}g_{\mu\nu}$.
$F_{\mu\nu}$ is the usual Yang-Mills field strength, here in the adjoint
representation of ${E_8}\otimes{E_8}$ or $SO(32)$, and $R$ is the usual
Ricci scalar, but $H_{\mu\nu\rho}$ is a generalized field strength for the
antisymmetric tensor field, modified by a Chern-Simons three-form so that,
in terms of forms, $H={\rm d}B-{\alpha' \over 30}{\rm tr}\left( AF-{1 \over
3}gA^3 \right)$. The equations of motion following from this action can be
written, after some rearrangement of the dilaton and curvature equations,
as
	\begin{eqn}								\label{equations of motion}
		\tb
			{\rm e}^{\phi/\alpha} {\nabla_\mu} \left(
				{\rm e}^{-\phi/\alpha} {g^{\mu\nu}} {\nabla_\nu}
				\left(\phi/\alpha\right)
				\right)
			+ {\smfr{2}{3}} {H^2} + {\frac{\alpha'}{30}} {\rm tr} {F^2}
			= 0,
		& \\ \tb
			{\rm e}^{\phi/\alpha} {\nabla_\mu}
				\left( {\rm e}^{-\phi/\alpha} {H^{\mu\nu\rho}} \right)
			= 0,
		& \\ \tb
			{\rm e}^{\phi/\alpha} {D_\mu}
				\left( {\rm e}^{-\phi/\alpha} {F^{\mu\nu}} \right)
			+ {H^{\nu\rho\sigma}} {F_{\rho\sigma}}
			= 0,
		& \\ \tb
			R_{\mu\nu} = {\kappa^2} S_{\mu\nu} =
			- {\nabla_\mu}{\nabla_\nu} {\left(\phi/\alpha\right)}
			+ {H_{\mu\rho\sigma}}{H_\nu^{\ \rho\sigma}}
			+ {\frac{\alpha'}{15}} {\rm tr} {F_{\mu\rho}}{F_\nu^{\ \rho}},
	\end{eqn}
where $\nabla_\mu$ is the spacetime-covariant derivative and $D_\mu$ is the
full spacetime- and gauge-covariant derivative.

The simplest string-like solution with gauge fields is that corresponding
to a straight, static macroscopic string with its worldsheet covering, say,
the $x^0$--$x^1$ plane, and which carries a $U(1)$ current, a source of
`electromagnetism'. It is spherically symmetric in the space transverse to
the worldsheet, and should be invariant under translations in either of the
worldsheet directions. This implies that the metric has the form
	\begin{equation}						\label{SS general metric}
		{\rm d}s^2 =
		{{\hat g}_{\alpha\beta}(r)} {\rm d}x^{\alpha} {\rm d}x^{\beta}
		+ {{\rm e}^{B(r)}} {\rm d}\vec{x} \cdot {\rm d}\vec{x},
	\end{equation}
where $r=\sqrt{{\vec x}\cdot{\vec x}}$ and $\alpha$ and $\beta$ are
worldsheet indices, taking values 0 or 1, whilst the vector notation refers
to the transverse space, subsequently also denoted by the index $i$ or $j$.
For $D>4$ the only other non-zero fields are $B_{01}$, $\phi$ and $A_0$ and
$A_1$, and again these depend only on the variable $r$. The gauge fields
are normalized by the following convention: for $SO(32)$, and in terms of
the $SO(16)$ subgroup of ${E_8}\otimes{E_8}$, a trace in the fundamental
representation ${\rm tr_{\rm fund}}$ is related to a trace in the adjoint
representation by ${\rm tr_{\rm fund}}={\rm tr_{\rm adj}}$, so since it is
natural to normalize the $U(1)$ fields to match the fundamental
representation, the electromagnetic kinetic term is taken to be simply
${1\over2}{\alpha'}{F^2}$.

This ansatz is partially solved by imposing the condition, applicable when
$D=10$, that the solution preserves half of the supersymmetries of the
theory, namely those generated by a supersymmetry parameter of negative
chirality in the transverse space, or equivalently of negative chirality on
the worldsheet. Since the ansatz is purely bosonic, the infinitesimal
supersymmetry variations of the bosonic fields are necessarily zero, and
only the variations of the fermionic fields need be considered. They are
	\begin{eqn}								\label{fermion susy transf}
		{\delta_\epsilon} \lambda  \eq  {\sqrt{2}\over4\kappa} \left[
				- {\gamma^\mu}{\partial_\mu}{\phi}
				+ {\smfr{1}{6}}{H_{\mu\nu\rho}}{\gamma^{\mu\nu\rho}}
				\right] {\epsilon}, \\
		{\delta_\epsilon} {\psi_\mu}  \eq  {1\over\kappa} \left[
				{\partial_\mu}
				+ {\smfr{1}{4}} \left(
					{\omega_\mu^{\ mn}} - {H_\mu^{\ mn}}
					\right) {\Gamma_{mn}}
				\right] {\epsilon}, \\
		{\delta_\epsilon} {\chi}  \eq
				-{1\over4g} {F_{\mu\nu}} {\gamma^{\mu\nu}} {\epsilon},
	\end{eqn}
where Greek indices refer to the spacetime-coordinate basis for the tangent
space and Roman indices to an orthonormal basis, the two of which are
related by zehnbeins $e_\mu^m$, while $\omega_\mu^{\ mn}$ is the
corresponding spin connection, and where $\Gamma^m$ are the ten-dimensional
Dirac matrices satisfying $\left\{ \Gamma^m,\Gamma^n \right\}=2\eta^{mn}$,
$\gamma^\mu={e^\mu_m}{\Gamma^m}$ and $\gamma^{{\mu_1}\cdots{\mu_n}}$ is the
antisymmetrized product ${\gamma^{[\mu_1}} \cdots {\gamma^{\mu_n]}}$ with
unit weight, and finally $g$ is the gauge coupling, proportional to
$\kappa/\sqrt{\alpha'}$ for the heterotic string.

The supersymmetry condition implies that the supersymmetry variations are
zero for any spinor parameter of the form $\epsilon={\rm e}^\sigma
\epsilon_0$ where $\sigma$ is some unknown function on the transverse
space, and $\epsilon_0$ is a constant spinor of negative chirality on the
worldsheet, that is ${\Gamma^0}{\Gamma^1}{\epsilon_0}=-\epsilon_0$. Upon
substituting the ansatz and using this form of $\epsilon$, each of the
supersymmetry variations has a factor of the form $\left(a+b
\Gamma^0\Gamma^1 \right)\epsilon_0$. If the supersymmetry condition holds,
the $a$ and $b$ in each expression must be equal, giving the relations,
	\begin{eqn}								\label{susy constraints}
		{\sqrt{-\hat g}} {\,} {\partial_i} {\phi}
			\eq -{\smfr{1}{2}} {\epsilon}^{\alpha\beta}
				{H_{i\alpha\beta}}, \\
		{H_{i\alpha\beta}}
			\eq {\smfr{1}{2}} {\partial_i} {\sqrt{-\hat g}} {\,}
				{\epsilon}_{\alpha\beta}, \\
		{\sqrt{-\hat g}} {\,} {\hat g}^{\alpha\beta} {F_{i\beta}}
			\eq {\epsilon}^{\alpha\beta} {F_{i\beta}}, \\
		{\sqrt{-\hat g}} {\partial_i} A
			\eq -{\smfr{1}{2}} \left(
				{\sqrt{-\hat g}}{\,}{\omega_i^{\ 01}}
				+ H_{i01} \right), \\
		\cen{
			\begin{array}{ccccc}\ds
				{\hat g}_{00} =
					-{\lambda}^{-1} \left( \sqrt{-\hat g}
					+ {\hat g}_{01} \right),
			\tb\tb
				{\hat g}_{11} =
					{\lambda} \left( \sqrt{-\hat g}
					- {\hat g}_{01} \right),
			\tb\tb
				{\rm e}^B = \mu,
			\end{array}
			}
	\end{eqn}
where $\epsilon^{\alpha\beta}$ is an antisymmetric matrix with
$\epsilon^{01}=1$, and $\lambda$ and $\mu$ are constants. By choosing a
particular coordinate system, for which the metric is asymptotically
Minkowski, the constants can be set to unity, so that by further choosing a
particular gauge for $B_{\mu\nu}$ and taking $\phi$ to be asymptotically
zero, the ansatz, now generalized to $D$-dimensional spacetime, becomes
	\begin{eqn}								\label{ansatz with susy}
		B \eq 0, \\
		\cen{
			{\hat g}_{\alpha\beta}
			= {\rm e}^{E(r)} \left(
				\begin{array}{cc}\ds
					-\left( 1+C(r) \right)  \tb C(r) \nl
					C(r)  \tb 1-C(r)
				\end{array}
				\right),
			} \\
		{\rm e}^{\phi/\alpha} = -2 B_{01} \eq {\rm e}^{E(r)}, \\
		A_0 = - A_1 \eq M(r).
	\end{eqn}
Furthermore $\sigma={1\over4}E$ and the Chern-Simons three-form is zero, so
that $H=dB$. As for the ansatz in~\cite{sustrs & solitons}, the assumption
is that the only dependence on $D$ is through $\alpha$ in the dilaton
relation. In fact, apart from the presence of gauge fields and the
off-diagonal terms in $\hat g$, (\ref{ansatz with susy})~is identical to
the ansatz of~\cite{sustrs & solitons}; furthermore, for $D=10$, it is the
most general ansatz consistent with the spacetime symmetries and
supersymmetries of the simplest string-like configuration.

The solution is completed by substitution into the equations of
motion~(\ref{equations of motion}).
To allow for multiple-string solutions, $E$, $C$ and $M$ are now taken to
be functions of all the transverse coordinates $x^i$, rather than of $r$
alone. The ansatz is consistent; the antisymmetric tensor and dilaton
equations give
	\begin{equation}						\label{equation for E}
		{\eta^{ij}} {\partial_i\partial_j} {\rm e}^{-E} = 0,
	\end{equation}
and the equation for the gauge field gives
	\begin{equation}						\label{equation for A}
		{\eta^{ij}} ( \partial_i\partial_j M
			- 2 \partial_i E \partial_j M ) = 0,
	\end{equation}
whilst the graviton equation gives~(\ref{equation for E}) again, together
with
	\begin{equation}						\label{equation for C}
		{\eta^{ij}} ( \partial_i\partial_j C
			- {2\alpha'} {\rm e}^{-E} \partial_i M \partial_j M )
			= 0.
	\end{equation}
Finally the substitutions $M={{\rm e}^E}{N}$ and $C=R+{\alpha'}{{\rm
e}^E}{M^2}$ give the Laplace equations:
	\begin{equation}						\label{3 Laplace equations}
		{\eta^{ij}} {\partial_i\partial_j} {\rm e}^{-E}
		= {\eta^{ij}} {\partial_i\partial_j} N
		= {\eta^{ij}} {\partial_i\partial_j} R = 0.
	\end{equation}

The solution corresponding to a single string, with the metric
asymptotically flat and the gauge choice that $A_\mu$ is asymptotically
zero, has
	\begin{equation}						\label{single-str E,N,R}
		\begin{array}{ccccc}\ds
			{\rm e}^{-E} = 1 + 2m{\kappa^2} \Lambda,  \tb\tb
			N = q \Lambda, \tb\tb
			R = -2p{\kappa^2} \Lambda,
		\end{array}
	\end{equation}
where $m$, $q$ and $p$ are constants and
	\begin{equation}						\label{Lambda}
		\Lambda(r) = \left\{
			\begin{array}{ll}\ds
				{1 \over (D-4){\omega_{D-3}}{r^{D-4}}}  &
					\mbox{if $D>4$} \nl
				-{1 \over 2\pi} \log r  &  \mbox{if $D=4$}
			\end{array} \right.,
	\end{equation}
with $\omega_{D-3}$ the volume of the ($D-3$)-sphere, so that the fields
are given by
	\begin{eqn}								\label{single-str fields}
		{\rm d}{s^2} \eq {1 \over 1 + 2m{\kappa^2}\Lambda} \left[
					\eta_{\alpha\beta}
					+ \left( 2p{\kappa^2}\Lambda
						- {{\alpha'q^2\Lambda^2}
							\over{1+2m{\kappa^2}\Lambda}}
						\right)
						\left(
							\begin{array}{rr}
								1  &  -1  \\ -1  &  1
							\end{array}
							\right)_{\!\!\alpha\beta}
					\right] {\rm d}{x^\alpha}{\rm d}{x^\beta}
				+ {\rm d}{\vec x}{\cdot}{\rm d}{\vec x}, \\
		{\rm e}^{-\phi/\alpha} \eq 1 + 2m{\kappa^2}\Lambda, \\
		F_{i0} = -F_{i1} \eq
				-\left[{x^i/r}\over{\omega_{D-3}}{r^{D-3}}\right]
				{q \over \left(1+2m{\kappa^2}\Lambda\right)^2 }, \\
		H_{i01} \eq
				-\left[{x^i/r}\over{2\omega_{D-3}}{r^{D-3}}\right]
				{2m{\kappa^2} \over \left(1+2m{\kappa^2}\Lambda\right)^2 }.
	\end{eqn}
The asymptotic forms of the gauge and antisymmetric tensor fields show the
electric charge per unit length on the string equals the electric current
along the string, both given by $q$, and that the charge per unit length
acting as a source for $B_{\mu\nu}$, using a normalization compatible
with~\cite{sustrs & solitons}, is given by $m$. The solution is unchanged
under a Lorentz boost along $x^1$, except that the charge $q$ is rescaled.
This is as might be expected since the electric current is a null vector
$(q,-q)$ on the worldsheet.

As in~\cite{sustrs & solitons}, the ADM prescription is used to define the
mass and momentum per unit length of the solution, though with the physical
metric $\tilde g$, since this gives the usual form for the curvature term
in the action. Far away from the string, the fields are small enough to use
the linearized Einstein equations, with the metric written in terms of
$h_{\mu\nu}={\tilde g}_{\mu\nu}-\eta_{\mu\nu}$ which is taken to be small.
The total energy-momentum tensor $\theta_{\mu\nu}$, which includes a
contribution interpreted as the energy-momentum in the gravitational field,
is then defined as
	\begin{equation}						\label{total e-m tensor}
		{\kappa^2} \theta_{\mu\nu}
		= {R^{(1)}}_{\mu\nu}
			- {\smfr{1}{2}} \eta_{\mu\nu} {R^{(1)}}_{\rho}^{\rho},
	\end{equation}
where ${R^{(1)}}_{\mu\nu}$ is the linearized Ricci tensor,
	\begin{equation}						\label{linearised R}
		{R^{(1)}}_{\mu\nu} = {\smfr{1}{2}} \left(
			{\partial_\rho}{\partial_\nu}{h^\rho_\mu}
			+ {\partial_\rho}{\partial_\mu}{h^\rho_\nu}
			- {\partial_\mu}{\partial_\nu}{h^\rho_\rho}
			- {\partial^\rho}{\partial_\rho}{h_{\mu\nu}}
			\right),
	\end{equation}
where all indices are raised and lowered with $\eta_{\mu\nu}$. For the
single string solution
	\begin{equation}						\label{single str e-m tensor}
		\theta_{\alpha\beta} = {1\over2\kappa^2} {\partial^i} \left[
			-{\partial_i}{h_{\alpha\beta}}
			+ {\eta_{\alpha\beta}} \left(
				{\partial_i}{h^\gamma_\gamma} + {\partial_i}{h^j_j} -
				{\partial_j}{h^j_i}
				\right)
			\right],
	\end{equation}
and integrating over the transverse space defines
	\begin{eqn}								\label{single str mass/mom}
		\Theta_{\alpha\beta}  \eq
			\int_{}^{} {\!} {\rm d}^{D-2}x {\,}
			{\theta_{\alpha\beta}} \\
		\eq  {1\over2\kappa^2} \int_{S^\infty_{D-3}}^{}
			{{\rm d}\over{\rm d}r} \left[
				- {h_{\alpha\beta}}
				+ {\eta_{\alpha\beta}} \left(
					{h^\gamma_\gamma} + (D-3)h_{ii} \right)
				\right]
			r^{D-3} {\rm d}\Omega_{D-3} \\
		\eq  \left(
			\begin{array}{cc}\ds
			m + p  \tb  - p \nl
			- p  \tb  - m + p
			\end{array} \right),
	\end{eqn}
where $S^\infty_{D-3}$ is the ($D-3$)-sphere at spatial infinity in the
transverse space, and there is no sum over the repeated $i$ index. This
suggests interpreting the solution as a string with mass per unit length
$m$, equal as usual to the tension, but with, in addition, momentum per
unit length $p$ running down the string in the $-x^1$ direction at the
speed of light.

How does this solution relate to the fundamental strings discussed in
section~\ref{susy stuff}? Exciting the gauge sector of the heterotic string
introduces left-moving, current-carrying modes. However, the momentum and
charge  per unit length of these modes should be related, so the freedom to
choose $p$ and $q$ independently in this solution probably corresponds to
the freedom to excite additional neutral modes, which are not present for a
heterotic string. To identify properly this solution as representing the
fields around such a macroscopic string would require matching the field
singularities with string source terms from, say, the sigma model action
for the heterotic string, as in~\cite{sustrs & solitons}, with the
expectation that this would then relate $p$ and $q$.

Finally, because of the linearity of Laplace's equation, multiple-string
solutions are also clearly allowed. To be explicit,
	\begin{eqn}								\label{multi-str solution}
		{\rm e}^{-E}  \eq  1 + {2{\kappa^2}} \sum^{}_k {m_k}
			{\Lambda( \left|{\vec x}-{\vec x}_k\right| )}, \\
		N  \eq  {2{\kappa^2}} \sum^{}_k {q_k}
			{\Lambda( \left|{\vec x}-{\vec x}_k\right| )}, \\
		R  \eq  -{2{\kappa^2}} \sum^{}_k {p_k}
			{\Lambda( \left|{\vec x}-{\vec x}_k\right| )},
	\end{eqn}
gives the solution for a collection of parallel strings labelled by $k$, so
that a given string has mass per unit length $m_k$, carries charge per unit
length $q_k$ and momentum per unit length $p_k$, and is located at position
${\vec x}_k$ in the transverse space.

\DWsection{A charged black string solution}	\label{another solution}

Starting with the same bosonic action, Horowitz and Strominger~\cite{black
strs} have found a family of singular, string-like solutions, now each with
an event horizon, for which the solution of Dabholkar {\em
et~al.}~\cite{sustrs & solitons} is the extremal case where the event
horizon coincides with the singularity. Given that the solution of the
section \ref{a solution} is simply a generalization of the solution of
\cite{sustrs & solitons}, similarly without a horizon, is there then a
corresponding family of `black string' solutions, each now also a source of
gauge fields, for which it is the extremal case?

For a solution to have a horizon, the sigma-model metric on the transverse
space cannot be flat, so by the discussion in the first part of
section~\ref{a solution}, black strings cannot preserve half the
supersymmetries. (Rather the expectation is that only the particular
extremal case, where the mass per unit length and the charge per unit
length which acts as the source of the antisymmetric tensor field are set
equal, saturates the relevant Bogomol'nyi inequality and allows the
addition symmetry.) Retaining, however, the other symmetries of a straight,
static, string-like solution, the fields about a black string have the
general form:
	\begin{eqn}								\label{black str ansatz}
		\tb
			{\rm d}s^2 = {{\hat g}_{\alpha\beta}(r)}
				{\rm d}x^{\alpha} {\rm d}x^{\beta}
			+ {{\rm e}^{B(r)}} {\rm d}r^2 + r^2 {\rm d}\Omega^2_{D-3},
		& \\ \tb
			\begin{array}{ccccc}\ds
				\phi/\alpha = E(r),  \tb\tb
				H_{rtx} = H(r),  \tb\tb
				A_\alpha = A_\alpha(r),
			\end{array}
	\end{eqn}
where ${\rm d}\Omega_{D-3}$ is the volume element on the ($D-3$)-sphere and
otherwise the notation follows that of section~\ref{a solution}.

Since there are no other constraints on the form of the solution, all that
is left is to do is to substitute the ansatz in the equations of
motion~(\ref{equations of motion}), though with the physical conditions
that asymptotically the metric should be flat and the fields should go to
zero, that the spacetime should have a singularity surrounded by a horizon
and that the fields must be finite everywhere away from the singularity.
The algebra is slightly simplified by changing variables from $r$ to
$\Lambda$ and introducing
	\begin{equation}						\label{hat variables}
		\begin{array}{ccc}\ds
			{\hat H} = -2 (D-4) {\omega_{D-3}} r^{D-3} H,  \tb\tb
			{\hat F}_\alpha = - (D-4) {\omega_{D-3}} r^{D-3}
				{{\rm d}{A_\alpha}\over{{\rm d}r}}
			= {\dot {A_\alpha}},
		\end{array}
	\end{equation}
where the dot represents differentiation with respect to $\Lambda$. In
particular, the equations for $\phi/\alpha$, ${\hat
g}^{\alpha\beta}R_{\alpha\beta}$, and the components of $R_{\mu\nu}$
parallel to the $(D-3)$-sphere are given by
	\begin{eqn}								\label{A,E & B equations}
		\tb
			\ddot{E} + \left(
				{\smfr{1}{2}} \dot{\log{\left|{\hat g}\right|}}
				- {\smfr{1}{2}} \dot{B} - \dot{E}
				\right) \dot{E}
			- \left|{\hat g}\right|^{-1} {\hat H}^2
			+ {\alpha'} {\hat F}^{\rm T} {\hat g}^{-1} {\hat F} = 0,
		& \\ \tb
			{\smfr{1}{2}}{\ddot{\log{\left|{\hat g}\right|}}} + \left(
				{\smfr{1}{2}} \dot{\log{\left|{\hat g}\right|}}
				- {\smfr{1}{2}} \dot{B} + \dot{E}
				\right) {\smfr{1}{2}}{\dot{\log{\left|{\hat g}\right|}}}
			- \left|{\hat g}\right|^{-1} {\hat H}^2
			+ {\alpha'} {\hat F}^{\rm T} {\hat g}^{-1} {\hat F} = 0,
		& \\ \tb
		{\Lambda} \left( {\smfr{1}{2}} \dot{\log{\left|{\hat g}\right|}}
				- {\smfr{1}{2}} \dot{B} + \dot{E} \right)
			+ {\rm e}^B - 1 = 0,
	\end{eqn}
respectively, where $\left|{\hat g}\right|=-{\rm det}{{\hat
g}_{\alpha\beta}}$ and a matrix notation is used so that there is an
implied summation over $\alpha=0,1$ in expressions such as ${\hat F}^{\rm
T}{\hat g}^{-1}{\hat F}$, with the superscript~${}^{\rm T}$ representing
transposition and ${\hat g}^{-1}={\hat g}^{\alpha\beta}$.

This set of equations can be partially solved. Taking the difference of the
first two equations, together with the third gives a pair of equations in
only the fields $C=\log{\left|{\hat g}\right|}-2E$ and $B$,
	\begin{equation}						\label{C & B equation}
		\begin{array}{ccc}\ds
			2 \ddot{C} + \left( \dot{C} - \dot{B} \right) \dot{C} = 0,
			\tb\tb
			2 \left( {\rm e}^B - 1 \right)
				+ {\Lambda} \left( \dot{C} - \dot{B} \right) = 0,
		\end{array}
	\end{equation}
which have the physical solution
	\begin{equation}						\label{B solution}
		\begin{array}{ccc}\ds
			{\rm e}^{-B} = 1 - a{\Lambda}, \tb\tb
			{\smfr{1}{2}}\log{\left|{\hat g}\right|} = E - {\smfr{1}{2}}B,
		\end{array}
	\end{equation}
where $a$ is a constant. The remaining unsolved equation, in terms of $E$,
is then
	\begin{equation}						\label{E equation}
		 \ddot{E} - \dot{B} \dot{E}
			- \left|{\hat g}\right|^{-1} {\hat H}^2
			+ {\alpha'} {\hat F}^{\rm T} {\hat g}^{-1} {\hat F} = 0.
	\end{equation}
Furthermore, the equation for $\hat H$,
	\begin{equation}						\label{H equation}
		\dot{\hat H}
		+ \left( {1\over2}\dot{{\rm log}\left|{\hat g}\right|}
			- {1\over2}\dot{B} + \dot{E} \right) {\hat H}
		= 0,
	\end{equation}
then has the solution, with arbitrary constant $c$,
	\begin{equation}						\label{H solution}
		{\hat H} = c {\,} {\rm e}^{2E},
	\end{equation}
and taken together, the expressions~(\ref{B solution}) and~(\ref{H
solution}) are unchanged from those for the black string without gauge
fields of~\cite{black strs}, though here the form of $E$ is as yet unknown.

The remaining unsolved equations of motion are for $F_{r\alpha}$,
$R_{\alpha\beta}$ and $R_{rr}$ and in matrix notation are given by
	\begin{num}								\label{F & g equations}
	\begin{eqnarray}
		\tb									\label{F equation}
			{\dot{\hat F}} - {\dot B}{\dot{\hat F}}
				+ {\hat g}{\dot{\hat g}}^{-1}{\hat F}
				+ c{\,}{\rm e}^B{\hat g}{\epsilon}{\hat F}
				= 0,
		& \\ \tb							\label{g equation}
			{\ddot{\hat g}} - {\dot B}{\dot{\hat g}}
				+ {\dot{\hat g}}{\dot{\hat g}}^{-1}{\hat g}
				- c^2{\rm e}^{2E+B} + 2{\alpha'}{\hat F}{\hat F}^{\rm T}
				= 0,
		& \\ \tb							\label{constraint}
			{\smfr{1}{2}}{\dot B}^2
				+ {\smfr{1}{2}}{\rm tr}{\dot{\hat g}}{\dot{\hat g}}^{-1}
				+ c^2{\rm e}^{2E+B}
				- 2{\alpha'}{\hat F}^{\rm T}{\hat g}^{-1}{\hat F}
				= 0,
	\end{eqnarray}
	\end{num}
respectively, where, as before, $\epsilon$ is an antisymmetric matrix with
$\epsilon^{01}=1$. Clearly the unsolved equation for $E$ is included, as
can be seen by contracting~(\ref{g equation}) with ${\hat g}^{\alpha\beta}$
and using the relation~(\ref{B solution}) between $\log{\left|{\hat
g}\right|}$ and $E$. Thus, what remains is to solve equations~(\ref{F & g
equations}) for the fields $\hat g$ and $\hat F$. However the equations are
nonlinear and coupled, and it appears difficult to make further progress.

The step to finding the full solution is to notice that the unknown fields
enter the equations for $E$~(\ref{E equation}) and
$R_{rr}$~(\ref{constraint}) in only particular combinations --- as $E =
{1\over2}\log{\left|{\hat g}\right|} + {1\over2}B$, ${\hat F}^{\rm T}{\hat
g}^{-1}{\hat F}$ and ${\rm tr}{\dot{\hat g}}{\dot{\hat g}}^{-1}$.
Furthermore if~(\ref{F equation}) is contracted with ${{\hat F}_\beta}{\hat
g}^{\beta\alpha}$ and~(\ref{g equation}) with ${\dot{\hat
g}}^{\alpha\beta}$ and ${{\hat F}^\alpha}{{\hat F}^\beta}$, the same
combinations of fields appear, though together with two new combinations,
${\hat F}^{\rm T}{\dot{\hat g}}^{-1}{\hat F}$ and ${\hat F}^{\rm T}{\hat
g}^{-1}{\dot{\hat F}}$. In fact, a new set of variables can be introduced,
	\begin{eqn}								\label{alpha definitions}
		\tb
		\begin{array}{ccc}\ds
			E = {\smfr{1}{2}} \log{\left|{\hat g}\right|} + {\smfr{1}{2}}B,
			\tb\tb
			{\alpha} = {\smfr{1}{2}}{\,}{\rm e}^{-4E}
				{\rm tr} {\dot{\hat g}}\epsilon {\dot{\hat g}}\epsilon,
		\end{array}
		& \\ \tb
		\begin{array}{ccccc}\ds
			{\beta} = {\alpha'}{\,}{\rm e}^{-4E}
				{\hat F}^T \epsilon{\hat g}\epsilon {\hat F}, \tb\tb
			{\gamma} = {\alpha'}{\,}{\rm e}^{-4E}
				{\hat F}^T \epsilon{\dot {\hat g}}\epsilon {\hat F}, \tb\tb
			{\delta} = {\alpha'}{\,}{\rm e}^{-4E}
				{\hat F}^T \epsilon{\ddot {\hat g}}\epsilon {\hat F},
		\end{array}
	\end{eqn}
which are related to the field combinations already mentioned, (it is
helpful to recall that, since ${\hat g}$ is symmetric, ${\hat
g}^{-1}=\left|{\hat g}\right|^{-2}\epsilon{\hat g}\epsilon$), and in terms
of which the equations for $E$~(\ref{E equation}) and
$R_{rr}$~(\ref{constraint}) become
	\begin{eqn}								\label{alpha equations 1}
		{\rm e}^{-2E-B} \left( {\ddot E} - {\dot B}{\dot E} \right)
			\eq - {\beta} + c^2, \\
		{\rm e}^{-2E-B} \left( {\dot E}^2 - {\dot B}{\dot E} \right)
			\eq - {\smfr{1}{2}}{\alpha} + {\beta} - {\smfr{1}{2}}c^2,
	\end{eqn}
whilst the contraction of (\ref{F equation}) with ${{\hat F}_\beta}{\hat
g}^{\beta\alpha}$ and (\ref{g equation}) with ${\dot{\hat
g}}^{\alpha\beta}$ and ${{\hat F}^\alpha}{{\hat F}^\beta}$ give
	\begin{eqn}								\label{alpha equations 2}
		\tb
			{\dot {\beta}} + {\gamma} = 0,
		& \\ \tb
			{\dot {\alpha}} + \left(
				{\alpha} - {c^2} \right) \left( 2{\dot E} - {\dot B} \right)
			+ 4{\gamma} = 0,
		& \\ \tb
			{\delta} - 2{\dot E}{\gamma} + \left( {\alpha} - {c^2} \right)
				{\rm e}^{2E+B} {\beta}
			= 0,
	\end{eqn}
respectively.

Eliminating $\alpha$, $\beta$ and $\gamma$ between (\ref{alpha equations
1}) and (\ref{alpha equations 2}) gives
	\begin{equation}						\label{E reduced equation}
		\ddot {\left({\rm e}^{-E}\right)} =
			{\rm constant} {\,} {\rm e}^{2B} {\rm e}^{-E}.
	\end{equation}
However, there is the physical condition that the fields ${\rm
e}^{-\phi/\alpha} = {\rm e}^{-E}$ and $H_{r01} = -{c{\,}{\rm
e}^{2E}}/{2{\omega_{D-3}}r^{D-3}}$ are non-singular on the horizon at ${\rm
e}^{-B} = 0$, which implies that the only physical solution of (\ref{E
reduced equation}) is
	\begin{equation}						\label{E solution}
		{\rm e}^{-E} = 1 + b {\Lambda},
	\end{equation}
in fact, as was the case in all previous solutions. Solving the remaining
equations gives the simple relations
	\begin{equation}						\label{alpha solutions}
		\begin{array}{lcl}\ds
			{\alpha} = {\smfr{1}{2}}{\rm e}^{-4E}
				{\rm tr}{\dot{\hat g}}\epsilon {\dot{\hat g}}\epsilon
			= {c^2},
		\tb\tb
			{\beta} = {\rm e}^{-4E}
				{\hat F}^T \epsilon{\hat g}\epsilon {\hat F}
			= {c^2} - \left( a + b \right) b,
		\nl
			{\gamma} = {\rm e}^{-4E}
				{\hat F}^T \epsilon{\dot {\hat g}}\epsilon {\hat F}
			= 0,
		\tb\tb
			{\delta} = {\rm e}^{-4E}
				{\hat F}^T \epsilon{\ddot {\hat g}}\epsilon {\hat F}
			= 0.
		\end{array}
	\end{equation}

It is then easy to solve for $\hat g$ and $\hat F$ in these relations, and
since, except in the case $\beta=0$, the equations (\ref{alpha equations
1}) and (\ref{alpha equations 2}) are equivalent to the original unsolved
equations for ${\hat g}$ and ${\hat F}$ (\ref{F & g equations}), doing so
gives the most general black string solution,
	\begin{eqn}								\label{black str solution}
		{\rm d}{s^2} \eq {1\over{1+b\Lambda}} \left(
				-1 + { { {{q_0}^2}a + \left({q_0}^2+{q_1}^2\right)b +
						2{q_0}{q_1}c }
					\over {{q_0}^2-{q_1}^2} } \Lambda
				- { { {\alpha'}{q_0}^2{\Lambda}^2 } \over {1+b\Lambda} }
				\right) \left({\rm d}{x^0}\right)^2
		\\ & &
			{} + {2\over{1+b\Lambda}} \left(
				{ { {q_0}{q_1}a + 2{q_0}{q_1}b + \left({q_0}^2+{q_1}^2\right)c }
					\over {{q_0}^2-{q_1}^2} } \Lambda
				- { { {\alpha'}{q_0}{q_1}{\Lambda}^2 } \over {1+b\Lambda} }
				\right) {\rm d}{x^0} {\rm d}{x^1}
		\\ & &
			{} + {1\over{1+b\Lambda}} \left(
				1 + { { {{q_1}^2}a + \left({q_0}^2+{q_1}^2\right)b +
							2{q_0}{q_1}c }
						\over {{q_0}^2 - {q_1}^2 }} \Lambda
				- { { {\alpha'}{q_1}^2{\Lambda}^2 } \over {1+b\Lambda} }
				\right) \left({\rm d}{x^1}\right)^2
		\\ & &
			{} + { {\rm d}{r^2} \over {1-a\Lambda} }
			+ {r^2} {\rm d}{\Omega_{D-3}^2}, \\
		{\rm e}^{-\phi/\alpha} \eq 1+b\Lambda, \\
		F_{r\alpha} \eq - \left[1\over{\omega_{D-3}}r^{D-3}\right]
					{ {q_\alpha} \over \left(1+b\Lambda\right)^2 }, \\
		H_{r01} \eq - \left[1\over{2\omega_{D-3}}r^{D-3}\right]
				{ c \over \left(1+b\Lambda\right)^2 }.
	\end{eqn}
The constants $a$, $b$, $c$, $q_0$ and $q_1$ are related by
	\begin{equation}						\label{a,b,c relation}
		{c^2} = {\alpha'} \left({q_0}^2-{q_1}^2\right)
			+ \left(a+b\right) b,
	\end{equation}
and from the asymptotic forms of the fields, $q_0$ is the electric charge
per unit length, $q_1$ is the electric current and c is the antisymmetric
tensor field charge per unit length. Again the Chern-Simons three-form
vanishes for this solution.

As was the case in section~\ref{a solution}, the solution is unchanged
under Lorentz boosts along the $x^1$ axis, except that $q_0$ and $q_1$ are
replaced throughout by the boosted pair
	\begin{equation}						\label{boosted q's}
		\left( \begin{array}{c}\ds  {q_0}' \nl  {q_1}'  \end{array} \right)
		= \left( \begin{array}{cc}\ds
				{\rm cosh}{\phi}  \tb  {\rm sinh}{\phi} \nl
				{\rm sinh}{\phi}  \tb  {\rm cosh}{\phi}
			\end{array} \right)
			\left(
				\begin{array}{c}\ds  {q_0} \nl  {q_1}  \end{array}
			\right),
	\end{equation}
where $\phi$ is the angle of the Lorentz boost. Thus if ${q_0}^2 >
{q_1}^2$, by boosting to a frame where ${q_0}'=0$, the solution can be
written in the simpler form:
	\begin{eqn}								\label{boosted black str}
		{\rm d}{s^2} \eq - \left({\rm d}{x^0}\right)^2
			- {{2c\Lambda} \over {1+b\Lambda}} {\rm d}{x^0} {\rm d}{x^1}
			+ {{1-a\Lambda-{c^2}{\Lambda^2}}
					\over \left(1+b\Lambda\right)^2}
				\left({\rm d}{x^1}\right)^2
			+ {{\rm d}{r^2} \over {1-a\Lambda}}
			+ {r^2} {\rm d}{\Omega_{D-3}^2}, \\
		{\rm e}^{-\phi/\alpha} \eq 1 + b \Lambda, \\
		F_{r0} \eq 0, \makebox[4\arraycolsep]{}
			F_{r1} = - \left[1\over{\omega_{D-3} r^{D-3}}\right]
					{{q_1}' \over \left(1+b\Lambda\right)^2 }, \\
		H_{r01} \eq - \left[1 \over {2\omega_{D-3} r^{D-3}}\right]
				{c \over \left(1+b\Lambda\right)^2 },
	\end{eqn}
where now ${c^2} + {\alpha'}{{q_1}'}^2 = \left( a + b \right) b$. (Of
course if ${q_0}^2>{q_1}^2$ then the solution can be written in a similar
form by boosting so ${q_1}'=0$.)

Finally the extremal limit should be identified, as well as the limit
giving the black string solution of Horowitz and Strominger~\cite{black
strs}. For the former, taking independently $a \to 0$, ${q_0} \to q$ and
${q_1} \to -q$ gives
	\begin{eqn}								\label{extremal limit}
		{\rm d}{s^2} \eq {1 \over {1+b\Lambda}} \left[
				\left( - 1 + {{q^2}\Lambda/b} \right)
					\left({\rm d}{x^0}\right)^2
				- \left(2{q^2}\Lambda/b\right) {\rm d}{x^0} {\rm d}{x^1}
				+ \left( 1 + {{q^2}\Lambda/b} \right)
					\left({\rm d}{x^1}\right)^2 \right]
			\\ & &
			{} + {\rm d}{r^2} + {r^2}{\Omega_{D-3}^2}, \\
		{\rm e}^{-\phi/\alpha} \eq 1 + b \Lambda, \\
		F_{r0} = - F_{r1} \eq - \left[1\over{\omega_{D-3}}r^{D-3}\right]
				{ q \over \left(1+b\Lambda\right)^2 }, \\
		H_{r01} \eq - \left[1\over2{\omega_{D-3}}r^{D-3}\right]
				{ b \over \left(1+ b\Lambda\right)^2 },
	\end{eqn}
which is the supersymmetric solution of section~\ref{a solution}, with the
identification $m=b/2{\kappa^2}$ and $p=\alpha'q^2/2b{\kappa^2}$. Note
however that it is not quite the general solution which has $p$ and $q$
arbitrary. This is because the black string solution was constrained by
physical conditions which no longer necessarily hold in the extremal case.
In particular, the condition that the fields are finite on the horizon is
no longer applicable when the horizon coincides with the singularity, and
it is this additional freedom which allows an extra parameter to enter the
extremal solution.

The black string solution without gauge fields of Horowitz and Strominger
is easily obtained by writing ${q_0}=\sqrt{a+b}{\,}\epsilon$ and
${q_1}=-\sqrt{b}{\,}\epsilon$ and taking $\epsilon \to 0$ and $D=10$. This
gives
	\begin{eqn}								\label{black str limit}
		{\rm d}{s^2} \eq
			- {{1-a\Lambda} \over {1+b\Lambda}} \left({\rm d}{x^0}\right)^2
			+ {\left({\rm d}{x^1}\right)^2 \over {1+b\Lambda}}
			+ {{\rm d}{r^2} \over {1-a\Lambda}}
			+ {r^2} {\rm d}{\Omega_7^2}, \\*
		{\rm e}^{-2\phi} \eq  1 + b \Lambda, \\*
		H_{r01} \eq - { 3\over{r^7} }
			{ c \over \left(1+b\Lambda\right)^2 },
	\end{eqn}
together with $c = \sqrt {\left(a+b\right)b}$ so that in terms of the
parameters of~\cite{black strs} $a = 6 {\omega_7} C$, $b = 6 {\omega_7}
{r_-^6}$ and $c = 2 {\omega_7} Q$.

\DWsection{Conclusion}						\label{conclusion}

Two new, charged, string-like solutions of the equations of motion for the
low-energy effective field theory of the heterotic string have been
described. The first was the most general solution which preserves one half
of the supersymmetries, which is found to have equal electric current and
electric charge per unit length, a mass per unit length equal to the
antisymmetric-tensor-field charge per unit length, as in~\cite{sustrs &
solitons}, and furthermore allows multi-string solutions. The second was
the most general solution with a horizon, a charged black string, for which
the former solution is identified as an extremal limit where the horizon
coincides with the singularity.

To complete the picture, some other limits of these solutions should
briefly be mentioned. Taking the gauge charges, $q_0$ and $q_1$,
arbitrarily to zero in the charged black string solution gives the black
string of Horowitz and Strominger~\cite{black strs}, though in general a
Lorentz-boosted version. For the supersymmetric solution, taking $q$ to
zero gives the solution of Dabholkar {\em et~al.}~\cite{sustrs & solitons},
though now modified to include momentum running down the string, a case
which is in fact the most general uncharged, horizonless, string-like
solution. Finally simply taking $a=0$ in the charged black string solution
gives a horizonless solution with arbitrary current and charge per unit
length, and which does not preserve half the supersymmetries. Thus, unlike
the uncharged case, for a charged string it is possible to have a solution
without a horizon which nonetheless is not supersymmetric.

Throughout, the correspondence between these solutions and fundamental
strings has been stressed. Section~\ref{susy stuff} demonstrated that any
macroscopic heterotic string which is a ground state of the right-moving
sector (a case which includes the stable states) preserves one half of the
supersymmetries, in the same manner as the supersymmetric solution. It
would be interesting to strengthen this correspondence and identify the
singularities of the solutions with suitable fundamental-string source
terms, as was done in~\cite{sustrs & solitons}, perhaps not only for the
supersymmetric case but in general, when the fundamental string is no
longer in the right-moving ground state.

\DWsubsectionstar{Acknowledgements}

I would like to thank Jeff Harvey for suggesting the questions addressed in
this paper and in addition, together with Jerome Gauntlett and Mike
Robinson, for useful discussions and comments. This work is supported in
part by the National Science Foundation, grant PHY 91-23780, and is
presented as a thesis to the Department of Physics, The University of
Chicago, in partial fulfillment of the requirements for the Ph.D. degree.

\DWsubsectionstar{Note}

During the process of writing this paper I was made aware of a preprint by
A. Sen~\cite{Sen}, who was able to obtain essentially the same solutions by
making a `twist' transformation on the known uncharged supersymmetric and
black string solutions.

\end{document}